\documentclass[aps,prl,twocolumn,superscriptaddress,showpacs]{revtex4}

\usepackage{graphicx}
\usepackage{dcolumn}
\usepackage{bm}

\begin{document}

\preprint{CLNS 07/2013}       
\preprint{CLEO 07-17}         

\title{\boldmath Determination of the Strong Phase in $D^0\to K^+\pi^-$ Using
Quantum-Correlated Measurements}

\author{J.~L.~Rosner}
\affiliation{Enrico Fermi Institute, University of
Chicago, Chicago, Illinois 60637, USA}
\author{J.~P.~Alexander}
\author{D.~G.~Cassel}
\author{J.~E.~Duboscq}
\author{R.~Ehrlich}
\author{L.~Fields}
\author{L.~Gibbons}
\author{R.~Gray}
\author{S.~W.~Gray}
\author{D.~L.~Hartill}
\author{B.~K.~Heltsley}
\author{D.~Hertz}
\author{C.~D.~Jones}
\author{J.~Kandaswamy}
\author{D.~L.~Kreinick}
\author{V.~E.~Kuznetsov}
\author{H.~Mahlke-Kr\"uger}
\author{D.~Mohapatra}
\author{P.~U.~E.~Onyisi}
\author{J.~R.~Patterson}
\author{D.~Peterson}
\author{D.~Riley}
\author{A.~Ryd}
\author{A.~J.~Sadoff}
\author{X.~Shi}
\author{S.~Stroiney}
\author{W.~M.~Sun}
\author{T.~Wilksen}
\affiliation{Cornell University, Ithaca, New York 14853, USA}
% ...
\collaboration{CLEO Collaboration}
\noaffiliation

\date{March 27, 2008}

\begin{abstract} 
We exploit the quantum coherence between pair-produced $D^0$ and
$\bar D^0$ in $\psi(3770)$ decays to study charm mixing, which is characterized
by the parameters $x$ and $y$, and to make a first
determination of the relative strong phase $\delta$ between $D^0\to K^+\pi^-$
and $\bar D^0\to K^+\pi^-$.  Using 281 ${\rm pb}^{-1}$
of $e^+e^-$ collision data collected with the
CLEO-c detector at $E_{\rm cm}=3.77$ GeV, as well as branching fraction
input and time-integrated measurements of 
$R_{\rm M}\equiv (x^2+y^2)/2$ and 
$R_{\rm WS}\equiv\Gamma(D^0\to K^+\pi^-)/\Gamma(\bar D^0\to K^+\pi^-)$
from other experiments, we find
$\cos\delta = 1.03^{+0.31}_{-0.17}\pm 0.06$, where the
uncertainties are statistical and systematic, respectively.
By further including
other mixing parameter measurements, we obtain an alternate
measurement of $\cos\delta = 1.10\pm 0.35\pm 0.07$,
as well as $x\sin\delta = (4.4^{+2.7}_{-1.8}\pm 2.9)\times 10^{-3}$
and $\delta = (22^{+11}_{-12}$$^{+9}_{-11})^\circ$.
\end{abstract}

\pacs{12.15.Ff,13.20.Fc,13.25.Ft,14.40.Lb}
\maketitle

The phenomenon of charm mixing is conventionally described by
two small parameters, $x\equiv (M_2-M_1)/\Gamma$ and
$y\equiv (\Gamma_2-\Gamma_1)/2\Gamma$, where $M_{1,2}$ and
$\Gamma_{1,2}$ are the masses and widths, respectively, of the $CP$-odd ($D_1$)
and $CP$-even ($D_2$) neutral $D$ meson mass eigenstates, and
$\Gamma\equiv(\Gamma_1+\Gamma_2)/2$.  Many previous
searches for charm mixing have used $D^0$ decay times
to attain first-order sensitivity to $y$.  Lifetimes of $D^0$ decays to $CP$
eigenstates determine $y$, while doubly Cabibbo-suppressed (DCS)
transitions probe $R_{\rm M}\equiv (x^2+y^2)/2$ and a mode-dependent quantity,
$y'$.  For the most widely used DCS mode, $D^0\to K^+\pi^-$,
$y'\equiv y\cos\delta-x\sin\delta$, where $-\delta$ is the relative phase
between the DCS amplitude and the corresponding Cabibbo-favored
$\bar D^0\to K^+\pi^-$ amplitude:
$\langle K^+\pi^-|D^0\rangle/\langle K^+\pi^-|\bar D^0\rangle\equiv r e^{-i\delta}$.  
We adopt a convention in which $\delta$ corresponds to a strong
phase, which vanishes in the SU(3) limit~\cite{Gronau:2001nr}.
To date, $\delta$ has not been measured, so measurements of $y$ and $y'$
have not been directly comparable.
The magnitude $r$ of the amplitude ratio is approximately 0.06.

In this Letter, we implement the method described in
Ref.~\cite{Asner:2005wf} for measuring $y$ and $\cos\delta$ using
quantum correlations at the $\psi(3770)$
resonance~\cite{Gronau:2001nr,theory}, where
$D^0\bar D^0$ pairs produced in $e^+e^-$ collisions are in a $C$-odd
eigenstate.  We extract these parameters from decay rates to 
single tags (ST), which are individually reconstructed $D^0$ or $\bar D^0$
candidates, and double tags (DT), which are events where both $D^0$ and
$\bar D^0$ are reconstructed.
$CP$ violation in
$D$ and $K$ decays are negligible second order effects.

To first order in $x$ and $y$, the rate $\Gamma_{D^0\bar D^0}(i,j)$ for
$C$-odd $D^0\bar D^0$ decay to final state $\{i,j\}$ follows from the
anti-symmetric amplitude ${\cal M}_{ij}$:
\begin{eqnarray}
\nonumber
\Gamma_{D^0\bar D^0}(i,j) &\propto& {\cal M}^2_{ij} =
	\left| A_i \bar A_j - \bar A_i A_j \right|^2 \\
\label{eq:rates}
&=& \left|\langle i|D_2\rangle\langle j|D_1\rangle -
        \langle i|D_1\rangle\langle j|D_2\rangle \right|^2,
\end{eqnarray}
where $A_i\equiv\langle i|D^0\rangle$,
$\bar A_i\equiv\langle i|\bar D^0\rangle$, and we have used
$|D_{\shortstack[c]{{\scriptsize 1} \\ {\scriptsize 2}}}\rangle=[|D^0\rangle\pm|\bar D^0\rangle]/\sqrt{2}$.
Using $S_\pm$ and $e^\pm$ to denote $CP\pm$ eigenstates and semileptonic
final states, respectively, these amplitudes are normalized such that
${\cal B}_{K^-\pi^+}\approx A_{K^-\pi^+}^2(1+ry\cos\delta+rx\sin\delta)$,
${\cal B}_{S_\pm}\approx A_{S_\pm}^2(1\mp y)$, and
${\cal B}_e\approx A_e^2$.
Quantum correlations affect neither the total $D^0\bar D^0$ rate (and hence
the number ${\cal N}$ of $D^0\bar D^0$ pairs produced) nor the ST rates.
DT final states
with pairs of $CP$ eigenstates, however, are affected maximally;
same-$CP$ $\{S_\pm,S_\pm\}$ states are forbidden, while opposite-$CP$
$\{S_+,S_-\}$
states are doubled in rate relative to uncorrelated decay.  In general,
the correlations introduce interference terms that can depend on
$y$ and $\delta$.

$D^0\bar D^0$ decay involving a final $CP$ eigenstate naturally selects the
$D_1 D_2$ basis.  As a result, the branching fraction for an associated
semileptonic decay probes $y$.  While the semileptonic decay width itself
does not depend on the $CP$ eigenvalue, the total width of the parent $D_1$
or $D_2$ meson does:
$\Gamma_{\shortstack[c]{{\scriptsize 1} \\ {\scriptsize 2}}}=\Gamma(1\mp y)$.
Thus, the $D_{\shortstack[c]{{\scriptsize 1} \\ {\scriptsize 2}}}$ 
semileptonic branching fraction is ${\cal B}_e/(1\mp y)$, and
the effective quantum-correlated $D^0\bar D^0$ branching fraction
(${\cal F}^{\rm cor}$) for a $\{S_\pm,e\}$ final state is
${\cal F}^{\rm cor}_{S_\pm,e}\approx 2{\cal B}_{S_\pm}{\cal B}_e(1\pm y)$,
where the factor of 2 arises from the sum of $e^+$ and $e^-$ rates.
When combined with estimates of ${\cal B}_e$ and ${\cal B}_{S_\pm}$ from
ST yields, external sources, and flavor-tagged semileptonic yields,
this equation allows $y$ to be determined.

If an $S_+$ and a $K^-\pi^+$ decay occur in the same event, then
the $K^-\pi^+$ was produced by a $D_1$, and ${\cal F}^{\rm cor}_{S_+, K\pi}$
is
\begin{eqnarray}
\nonumber
{\cal F}^{\rm cor}_{S_+, K\pi} &=&
|\langle S_+|D_2\rangle \langle K^-\pi^+|D_1\rangle|^2 \\
\nonumber
&=& A_{S_+}^2 |A_{K^-\pi^+} + \bar A_{K^-\pi^+}|^2 \\
\nonumber
&=& A_{S_+}^2 A_{K^-\pi^+}^2 |1+re^{-i\delta}|^2 \\
&\approx& {\cal B}_{S_+}{\cal B}_{K^-\pi^+}(1+R_{\rm WS}+2r\cos\delta+y),
\end{eqnarray}
where $R_{\rm WS}$ is the wrong-sign rate ratio, which depends on $x$ and $y$
because of the interference between DCS and mixing transitions:
$R_{\rm WS} \equiv \Gamma(\bar D^0\to K^-\pi^+)/\Gamma(D^0\to K^-\pi^+) = r^2 + ry' + R_{\rm M}$. 
Similarly, the $\{S_-,K\pi\}$ DT yield probes
${\cal B}_{S_-}{\cal B}_{K^-\pi^+}(1+R_{\rm WS}-2r\cos\delta-y)$, and
the asymmetry between these two DT yields gives $\cos\delta$, given
knowledge of ${\cal B}_{S_\pm}$, $r$, and $y$.

Table~\ref{tab:rates} shows ${\cal F}^{\rm cor}$ for all categories
of final states considered in this analysis: $K^\mp\pi^\pm$, $S_\pm$, and
$e^\pm$.
Comparison of ${\cal F}^{\rm cor}$ with the uncorrelated effective
branching fractions, ${\cal F}^{\rm unc}$, also given in
Table~\ref{tab:rates}, provides $r\cos\delta$, $y$, $r^2$, $x^2$, and
$rx\sin\delta$.
These five parameters are extracted by combining our ST and
DT yields with external branching fraction measurements in a least-squares
fit~\cite{Sun:2005ip}.  The external measurements, from
incoherently produced $D^0$ mesons, provide one measure of ${\cal B}_i$. The
ST event yields provide a second measure; since each event has one $D^0$ and
one $\bar D^0$, inclusive rates correspond to uncorrelated branching fractions.
The fit averages these estimates, and we extract updated ${\cal B}_i$.
Finally, the DT/ST comparison provides ${\cal N}$, so the fit requires no
knowledge of luminosity or $D^0\bar D^0$ production cross sections.

\begin{table}[htb]
\caption{Correlated ($C$-odd) and uncorrelated effective $D^0\bar D^0$
branching fractions, ${\cal F}^{\rm cor}$ and
${\cal F}^{\rm unc}$, to leading order in $x$, $y$, and $R_{\rm WS}$,
divided by ${\cal B}_i$
for ST modes $i$ (first section) and
${\cal B}_i{\cal B}_j$ for DT modes $\{i,j\}$ (second section).
Charge conjugate modes are implied.}
\label{tab:rates}
\begin{tabular}{ccc}
\hline\hline
Mode & Correlated & Uncorr. \\
\hline
$K^-\pi^+$ &
        $1+R_{\rm WS}$ &
        $1+R_{\rm WS}$ \\
$S_\pm$ & $2$ & $2$ \\
\hline
$K^-\pi^+,K^-\pi^+$ &
        $R_{\rm M}$ &
        $R_{\rm WS}$ \\
$K^-\pi^+,K^+\pi^-$ &
        $(1+R_{\rm WS})^2-4r\cos\delta(r\cos\delta+y)$ &
        $1+R_{\rm WS}^2$ \\
$K^-\pi^+,S_\pm$ &
        $1+ R_{\rm WS}\pm 2r\cos\delta\pm y$ &
        $1+R_{\rm WS}$ \\
$K^-\pi^+,e^-$ &
        $1-ry\cos\delta-rx\sin\delta$ &
        $1$ \\
$S_\pm,S_\pm$ & 0 & $1$ \\
$S_+,S_-$ &
        $4$ &
        $2$ \\
$S_\pm,e^-$ &
        $1\pm y$ &
        $1$ \\
\hline\hline
\end{tabular}
\end{table}

We analyze 281 ${\rm pb}^{-1}$ of $e^+e^-$
collision data produced by the Cornell Electron Storage Ring (CESR) at
$E_{\rm cm}=3.77$ GeV and collected with the CLEO-c detector, which is
described in detail elsewhere~\cite{cleodetector}.
We reconstruct the $D^0$ and $\bar D^0$ final states listed in
Table~\ref{tab:finalStates}, with $\pi^0/\eta\to\gamma\gamma$,
$\omega\to\pi^+\pi^-\pi^0$, and $K^0_S\to\pi^+\pi^-$.
Signal and background efficiencies, 
as well as probabilities for misreconstructing a produced signal decay in a
different signal mode (crossfeed), 
are determined from simulated events that are processed in a
fashion identical to data.

\begin{table}[htb]
\caption{$D$ final states reconstructed in this analysis.}
\label{tab:finalStates}
\begin{tabular}{cc}
\hline\hline
Type & Final States \\
\hline
Flavored & $K^-\pi^+$, $K^+\pi^-$ \\
$S_+$ & $K^+K^-$, $\pi^+\pi^-$, $K^0_S\pi^0\pi^0$, $K^0_L\pi^0$ \\
$S_-$ & $K^0_S\pi^0$, $K^0_S\eta$, $K^0_S\omega$ \\
$e^\pm$ & Inclusive $Xe^+\nu_e$, $Xe^-\bar\nu_e$ \\
\hline\hline
\end{tabular}
\end{table}

The $D$ candidate selection and yield determination procedures are described
in a companion article~\cite{tqcaprd} and are summarized below.
Hadronic final states without $K^0_L$ mesons are fully reconstructed
via two kinematic variables: the beam-constrained candidate mass,
$M \equiv\sqrt{E_0^2/c^4 - {\mathbf p}_D^2/c^2}$, where ${\mathbf p}_D$ is the
$D^0$ candidate momentum and $E_0$ is the beam energy, and
$\Delta E\equiv E_D - E_0$, where
$E_D$ is the sum of the $D^0$ candidate daughter energies.
We extract ST and DT yields from $M$ distributions using
unbinned maximum likelihood fits (ST) or by counting candidates in signal
and sideband regions (DT).

\begin{table}[tb]
\caption{ST and DT yields, efficiencies, and their statistical
uncertainties. For DT yields, we sum groups of modes and provide an
average efficiency for each group;
the number of modes in each group is given in parentheses.
Modes with asterisks are not included in the standard and extended fits.}
\label{tab-STDTYieldsAndEffs}
\begin{tabular}{lcc}
\hline\hline
Mode &  ~~Yield~~ & ~~Efficiency (\%) \\ \hline
$K^-\pi^+$ &
	$25374\pm 168$ & $64.70\pm 0.04$ \\
$K^+\pi^-$ &
	$25842\pm 169$ & $65.62\pm 0.04$ \\
$K^+K^-$ &
	$4740\pm 71$ & $57.25\pm 0.09$ \\
$\pi^+\pi^-$ &
	$2098\pm 60$ & $72.92\pm 0.13$ \\
$K^0_S\pi^0\pi^0$ &
	$2435\pm 74$ & $12.50\pm 0.06$ \\
$K^0_S\pi^0$ &
	$7523\pm 93$ & $29.73\pm 0.05$ \\
$K^0_S\eta$ &
	$1051\pm 43$ & $10.34\pm 0.06$ \\
$K^0_S\omega$ &
	$3239\pm 63$ & $12.48\pm 0.04$ \\
\hline
$K^\mp\pi^\pm, K^\mp\pi^\pm$ (2) & $4\pm 2$ & $40.2\pm 2.4$ \\
$K^-\pi^+, K^+\pi^-$ (1) & $600\pm 25$ & $41.1\pm 0.2$ \\
$K^\mp\pi^\pm, S_+$ (8) & $605\pm 25$ & $26.1\pm 0.1$ \\
$K^\mp\pi^\pm, S_-$ (6) & $243\pm 16$ & $12.3\pm 0.1$ \\
$K^\mp\pi^\pm, e^\mp$ (2) & $2346\pm 65$ & $45.6\pm 0.1$ \\
$S_+, S_+$ (9*) & $10\pm 6$ & $12.5\pm 0.6$ \\
$S_-, S_-$ (6*) & $2\pm 2$ & $3.9\pm 0.2$ \\
$S_+, S_-$ (12) & $242\pm 16$ & $7.7\pm 0.1$\\
$S_+, e^\mp$ (6) & $406\pm 44$ & $22.2\pm 0.1$ \\
$S_-, e^\mp$ (6) & $538\pm 40$ & $13.8\pm 0.1$ \\
\hline\hline
\end{tabular}
\end{table}

Because most $K^0_L$ mesons and neutrinos produced at CLEO-c are not
detected,
we only reconstruct modes with these particles in DTs, where the
other $D$ in the event is fully reconstructed.
Ref.~\cite{ksklpi} describes the missing mass technique used to identify
$K^0_L\pi^0$ candidates.
For semileptonic decays, we use inclusive, partial reconstruction to maximize
efficiency, demanding only that the electron be identified.
Electron identification utilizes a
multivariate discriminant~\cite{eid} that combines measurements from the
tracking
chambers, the electromagnetic calorimeter, and the ring imaging
\v{C}erenkov counter.

Table~\ref{tab-STDTYieldsAndEffs} gives yields and efficiencies for
8 ST modes and 58 DT modes, where the DT modes have been grouped into
categories.  Fifteen of
the DT modes are forbidden by $CP$ conservation and are not included in
the nominal fits.
In general, crossfeed among signal modes and backgrounds from other $D$ decays
are smaller than 1\%.  Modes with $K^0_S\pi^0\pi^0$ have approximately
3\% background, and yields for $\{K^\mp\pi^\pm, K^\mp\pi^\pm\}$ and
$\{S_\pm, S_\pm\}$ are consistent with being entirely from background.

External inputs to the standard fit include measurements of $R_{\rm M}$,
$R_{\rm WS}$, ${\cal B}_{K^-\pi^+}$, and ${\cal B}_{S_\pm}$, as well as an
independent ${\cal B}_{K^0_L\pi^0}$ from CLEO-c, as shown in
Table~\ref{tab:externalMeas1}.  $R_{\rm WS}$ is required to
constrain $r^2$, and thus, to convert $r\cos\delta$ and $rx\sin\delta$ to
$\cos\delta$ and $x\sin\delta$.
We also perform an extended fit that uses the external mixing
parameter measurements shown in Table~\ref{tab:externalMeas2}.
These fits incorporate the full covariance matrix for these inputs,
accounting for statistical overlap with the yields in this analysis.
Covariance matrices for the fits in Ref.~\cite{wskpi} have been provided by
the CLEO, Belle, and BABAR collaborations.

\begin{table}[htb]
\begin{center}
\caption{Averages of external measurements used in the standard and extended
fits. Charge-averaged $D^0$ branching fractions are denoted by final state.}
\label{tab:externalMeas1}
\begin{tabular}{lc}
\hline\hline
Parameter & Average \\
\hline
$R_{\rm WS}$ & $0.00409\pm 0.00022$~\cite{rws}  \\
$R_{\rm M}$ & $0.00017\pm 0.00039$~\cite{rm} \\
$K^-\pi^+$   & $0.0381\pm 0.0009$~\cite{pdg04} \\
$K^-K^+/K^-\pi^+$     & $0.1010\pm 0.0016$~\cite{pdg06} \\
$\pi^-\pi^+/K^-\pi^+$ & $0.0359\pm 0.0005$~\cite{pdg06} \\
$K^0_L\pi^0$  & $0.0097\pm 0.0003$~\cite{ksklpi} \\ 
$K^0_S\pi^0$ & $0.0115\pm 0.0012$~\cite{pdg04} \\
$K^0_S\eta$  & $0.00380\pm 0.00060$~\cite{pdg04} \\
$K^0_S\omega$  & $0.0130\pm 0.0030$~\cite{pdg04} \\
\hline\hline
\end{tabular}
\end{center}
\end{table}

\begin{table}[htb]
\begin{center}
\caption{Averages of external measurements used only in the extended fit.}
\label{tab:externalMeas2}
\begin{tabular}{lc}
\hline\hline
Parameter & Average \\
\hline
$y$    & $0.00662\pm 0.00211$~\cite{pdg06,ycpBelle,kspipi} \\
$x$    & $0.00811\pm 0.00334$~\cite{kspipi} \\
$r^2$  & $0.00339\pm 0.00012$~\cite{wskpi} \\
$y'$   & $0.0034\pm 0.0030$~\cite{wskpi} \\
$x'^2$ & $0.00006\pm 0.00018$~\cite{wskpi} \\
\hline\hline
\end{tabular}
\end{center}
\end{table}

Systematic uncertainties include those associated with efficiencies for
reconstructing
tracks, $K^0_S$ decays, $\pi^0$ decays, and for hadron identification
(see Refs.~\cite{dhadprd,tqcaprd}).
Other sources of efficiency uncertainty include:
$\Delta E$ requirements (0.5--5.5\%), $\eta$ reconstruction (4.0\%),
electron identification (1.0\%),
modeling of particle multiplicity and detector noise (0.1--1.3\%),
simulation of initial and final state radiation (0.5--1.2\%),
and modeling of resonant substructure in $K^0_S\pi^0\pi^0$ (0.7\%).
We also include additive
uncertainties of 0.0--0.9\% to account for variations of yields with 
fit function.

These systematic uncertainties are included in the
covariance matrix given to the fitter, which propagates them to the fit
parameters.
The other fit inputs determined in this analysis are
ST and DT yields and efficiencies, crossfeed probabilities,
background branching fractions and efficiencies, and statistical
uncertainties on all of these measurements.  Quantum correlations between
signal and background modes are accounted for using assumed values of
amplitude ratios and strong phases that are systematically varied and found
to have negligible effect.
Using a simulated $C$-odd $D^0\bar D^0$ sample 15 times the size of our data
sample, we validated our analysis technique by reproducing
the input branching fractions and mixing parameters with biases due to our
procedures that were less than one-half of the statistical errors on the data
and consistent with zero.

Table~\ref{tab:DataResults} shows the results of the data fits, excluding
the 15 same-$CP$ DT modes.  Our standard fit includes the measurements in
Table~\ref{tab:externalMeas1} but not Table~\ref{tab:externalMeas2}.  
In this fit, $x\sin\delta$ is not determined reliably, so we fix it to zero,
and the associated systematic
uncertainty is $\pm 0.03$ for $\cos\delta$ and negligible for all other
parameters.
We obtain a first measurement of $\cos\delta$, consistent with being at the
boundary of the physical region.
Our branching fraction results do not supersede other CLEO-c measurements.

\begin{table}[htb]
\caption{Results from the standard fit (with Table~\ref{tab:externalMeas1}
inputs) and the extended fit (with
Table~\ref{tab:externalMeas1}/\ref{tab:externalMeas2} inputs).  Uncertainties
are statistical and
systematic, respectively.  Charge-averaged $D^0$ branching fractions are
denoted by final state.}
\label{tab:DataResults}
\begin{tabular}{lcc}
\hline\hline
Parameter & Standard Fit & Extended Fit \\
\hline
${\cal N}$ $(10^6)$ &
	$1.042\pm 0.021\pm 0.010$ &
        $1.042\pm 0.021\pm 0.010$ \\
$y$ $(10^{-3})$ &
	$-45\pm 59\pm 15$ &
        $6.5\pm 0.2\pm 2.1$ \\
$r^2$ $(10^{-3})$ &
        $8.0\pm 6.8\pm 1.9$ &
        $3.44\pm 0.01\pm 0.09$ \\
$\cos\delta$ &
        $1.03\pm 0.19\pm 0.06$ &
        $1.10\pm 0.35\pm 0.07$ \\
$x^2$ $(10^{-3})$ &
        $-1.5\pm 3.6\pm 4.2$ &
        $0.06\pm 0.01\pm 0.05$ \\
$x\sin\delta$ $(10^{-3})$ &
	0 (fixed) &
	$4.4\pm 2.4\pm 2.9$ \\
$K^-\pi^+$ (\%) &
        $3.78\pm 0.05\pm 0.05$ &
        $3.78\pm 0.05\pm 0.05$ \\
$K^-K^+$ $(10^{-3})$ &
        $3.87\pm 0.06\pm 0.06$ &
        $3.88\pm 0.06\pm 0.06$ \\
$\pi^-\pi^+$ $(10^{-3})$ &
        $1.36\pm 0.02\pm 0.03$ &
        $1.36\pm 0.02\pm 0.03$ \\
$K^0_S\pi^0\pi^0$ $(10^{-3})$ &
        $8.34\pm 0.45\pm 0.42$ &
        $8.35\pm 0.44\pm 0.42$ \\
$K^0_S\pi^0$ (\%) &
        $1.14\pm 0.03\pm 0.03$ &
        $1.14\pm 0.03\pm 0.03$ \\
$K^0_S\eta$ $(10^{-3})$ &
        $4.42\pm 0.15\pm 0.28$ &
        $4.42\pm 0.15\pm 0.28$ \\
$K^0_S\omega$ (\%) &
        $1.12\pm 0.04\pm 0.05$ &
        $1.12\pm 0.04\pm 0.05$ \\
$X^- e^+\nu_e$ (\%) &
        $6.54\pm 0.17\pm 0.17$ &
        $6.59\pm 0.16\pm 0.16$ \\
$K^0_L\pi^0$ (\%) &
        $1.01\pm 0.03\pm 0.02$ &
        $1.01\pm 0.03\pm 0.02$ \\
\hline
$\chi^2_{\rm fit}$/ndof &
        30.1/46 &
        55.3/57 \\
\hline\hline
\end{tabular}
\end{table}

\begin{figure}[htb]
\includegraphics*[width=\linewidth]{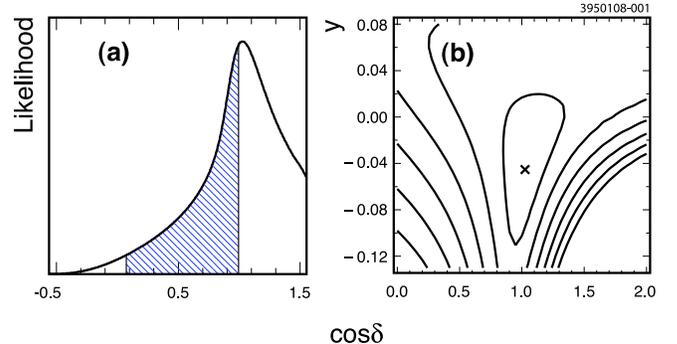}
\caption{Standard fit likelihood including both statistical and systematic
uncertainties for $\cos\delta$ (a), and simultaneous likelihood for
$\cos\delta$ and $y$ (b) shown as contours in increments of $1\sigma$, where
$\sigma=\sqrt{\Delta\chi^2}$.
The hatched region contains 95\% of the area in the physical region.}
\label{fig:standardFitLikelihoods}
\end{figure}

The likelihood curve for $\cos\delta$, shown in
Fig.~\ref{fig:standardFitLikelihoods}a, is
computed as ${\cal L}=e^{-(\chi^2-\chi^2_{\rm min})/2}$ at various fixed
values of $\cos\delta$.  It is highly non-Gaussian, so we assign asymmetric
uncertainties (which still do not fully capture the non-linearity) by finding
the values of $\cos\delta$ where $\Delta\chi^2=1$ to obtain
$\cos\delta = 1.03^{+0.31}_{-0.17}\pm 0.06$.
This non-linearity stems from the
use of $r\cos\delta$ to determine $\cos\delta$, which causes the
uncertainty on $\cos\delta$ to scale roughly like $1/r$.  Because $r^2$ is
obtained from
$R_{\rm WS}$, an upward shift in $y$ lowers the derived value of $r^2$
(for positive $r\cos\delta$), and the resultant uncertainty on $\cos\delta$
increases, as illustrated by Fig.~\ref{fig:standardFitLikelihoods}b.
For values of $\left|\cos\delta\right| < 1$, we also compute
${\cal L}$ as a function of $|\delta|$, and we integrate these
curves within the physical region to obtain 95\% confidence level (CL) limits
of $\cos\delta > 0.07$ and $|\delta| < 75^\circ$.

\begin{figure}[htb]
\includegraphics*[width=\linewidth]{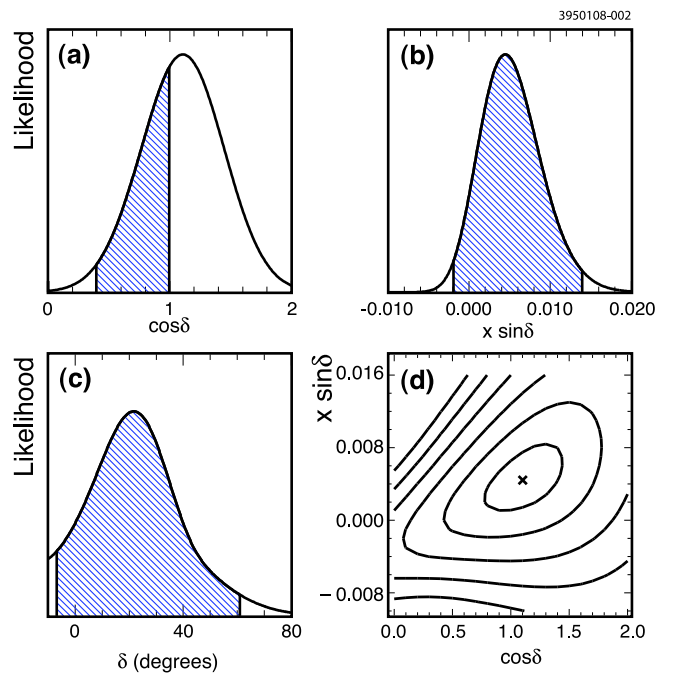}
\caption{Extended fit likelihood including both statistical and systematic
uncertainties for $\cos\delta$ (a), $x\sin\delta$ (b), $\delta$ (c), and
simultaneous likelihood for $\cos\delta$ and $x\sin\delta$ (d) shown as
contours in increments of $1\sigma$, where $\sigma=\sqrt{\Delta\chi^2}$.
The hatched
regions contain 95\% of the area in the physical regions.  
For $\delta$, the fit fails to converge beyond the limits of the plot.}
\label{fig:extendedFitLikelihoods}
\end{figure}

When combined with previous measurements of $y$ and $y'$, our measurement
of $\cos\delta$ also gives $x\sin\delta$.  Table~\ref{tab:DataResults}
shows the results of such an extended fit that includes external inputs from
both Table~\ref{tab:externalMeas1} and Table~\ref{tab:externalMeas2}.  
The resultant value of $y$ includes the
CLEO-c measurement from the standard fit, but the precision is
dominated by the external $y$ measurements.
The overall uncertainty on $\cos\delta$ increases to $\pm 0.36$ because of
the non-linearity discussed above.  However, unlike the standard fit,
the likelihood for $\cos\delta$ is nearly Gaussian, as shown in
Fig.~\ref{fig:extendedFitLikelihoods}a.
The correlation coefficient
between $\cos\delta$ and $x\sin\delta$ is 0.56, and we assign asymmetric
uncertainties of $x\sin\delta = (4.4^{+2.7}_{-1.8}\pm 2.9)\times 10^{-3}$.
By repeating the fit at various simultaneously fixed values of $\cos\delta$
and $\sin\delta$, we also determine
$\delta = (22^{+11}_{-12}$$^{+9}_{-11})^\circ$.
The corresponding 95\% CL intervals within the physical region
are $\cos\delta > 0.39$, $x\sin\delta\in [0.002, 0.014]$, and
$\delta\in [-7^\circ, +61^\circ]$.
Performing this extended fit with $y$, $x^2$, and $x\sin\delta$ fixed to zero
results in a change in $\chi^2$ of 25.1, or a significance of $5.0\sigma$.

By observing the change in $1/\sigma_y^2$ as
each fit input is removed, we identify the major contributors of
information on $y$ to be the $\{S_\pm, e\}$ yields (90\%) and $\{K\pi, e\}$
yields (10\%).
For $\cos\delta$, the $\{K\pi, S_\pm\}$ DT yields and the ST yields
simultaneously account for 100\%.
We also find that no single input or group of inputs exerts a pull larger
than $3\sigma$ on $\cos\delta$ or $y$.  Moreover, removing all external inputs
gives branching
fractions consistent with those in Table~\ref{tab:externalMeas1}.
Finally, if we determine $y$ only from $K^+K^-$ and $\pi^+\pi^-$ input, as in
previous direct measurements, the result is consistent with the value in
Table~\ref{tab:DataResults}.

We also allow for a $C$-even $D^0\bar D^0$ admixture in the initial state,
which is expected to be ${\cal O}(10^{-8})$~\cite{petrov}, by including
the 15 $\{S_\pm, S_\pm\}$ DT yields in the fit.
These modes limit the $C$-even component,
which can modify the other yields as described in
Ref.~\cite{Asner:2005wf}.  In both the standard and extended fits, we
find a $C$-even fraction consistent with zero
with an uncertainty of 2.4\%, and neither the fitted parameters nor their
uncertainties are shifted noticeably from the values in
Table~\ref{tab:DataResults}.

In summary, using 281 ${\rm pb}^{-1}$ of $e^+e^-$ collisions produced at the
$\psi(3770)$, we make a first determination of the strong phase $\delta$,
with $\cos\delta = 1.03^{+0.31}_{-0.17}\pm 0.06$.
By further including external
mixing parameter measurements in our analysis, we obtain an alternate
measurement of $\cos\delta = 1.10\pm 0.35\pm 0.07$, as well as
$x\sin\delta = (4.4^{+2.7}_{-1.8}\pm 2.9)\times 10^{-3}$
and $\delta = (22^{+11}_{-12}$$^{+9}_{-11})^\circ$.

We thank Alexey Petrov, William Lockman, Alan Schwartz, Bostjan Golob, and
Brian Petersen for helpful discussions.
We gratefully acknowledge the effort of the CESR staff 
in providing us with excellent luminosity and running conditions. 
This work was supported by 
the A.P.~Sloan Foundation, 
the National Science Foundation, 
the U.S. Department of Energy, and 
the Natural Sciences and Engineering Research Council of Canada.

\end{document}